\documentclass[%
 aip,
 superscriptaddress,
 amsmath,amssymb,
 reprint,%
 floatfix,
]{revtex4-1}
\usepackage{graphicx}
\usepackage{dcolumn}
\usepackage{bm}
\usepackage{siunitx}
\usepackage{xcolor}
\usepackage[utf8]{inputenc}
\usepackage[T1]{fontenc}
\usepackage{mathptmx}
\usepackage{etoolbox}
\usepackage[version=4]{mhchem} 
\usepackage{hyperref}
\makeatletter
\def\@email#1#2{%
 \endgroup
 \patchcmd{\titleblock@produce}
  {\frontmatter@RRAPformat}
  {\frontmatter@RRAPformat{\produce@RRAP{*#1\href{mailto:#2}{#2}}}\frontmatter@RRAPformat}
  {}{}
}%
\makeatother
\begin{document}

\preprint{}

\title[Laser-written waveguide-integrated coherent spins in diamond]{Laser-written waveguide-integrated coherent spins in diamond}
\author{Yanzhao Guo}
\affiliation{School of Engineering, Cardiff University, Queen’s Buildings, The Parade, Cardiff, CF24 3AA, United Kingdom}
 \affiliation{Translational Research Hub, Maindy Road, Cardiff, CF24 4HQ, United Kingdom}
 \author{John P. Hadden}
\affiliation{School of Engineering, Cardiff University, Queen’s Buildings, The Parade, Cardiff, CF24 3AA, United Kingdom}
\affiliation{Translational Research Hub, Maindy Road, Cardiff, CF24 4HQ, United Kingdom} 
\author{Federico Gorrini}
\affiliation{Department of Molecular Biotechnology and Health Sciences, University of Torino, Via Nizza 52, 10126 Torino, Italy}
\affiliation{Dipartimento Scienza Applicata e Tecnologia, Politecnico di Torino, Corso Duca degli Abruzzi 24, 10129, Torino, Italy}\author{Giulio Coccia}
\affiliation{Department of Physics, Politecnico di Milano, Piazza Leonardo da Vinci, 32, 20133 Milano, Italy}
\author{Vibhav Bharadwaj}
\affiliation{Institute for Photonics and Nanotechnologies (IFN-CNR), Piazza Leonardo da Vinci, 32, 20133 Milano, Italy}
\affiliation{Institute for Quantum Optics, Ulm University, D-89081 Ulm, Germany}
\affiliation{Department of Physics, Indian Institute of Technology, Guwahati, Assam, India 781039}
\author{Vinaya Kumar Kavatamane}
\affiliation{Institute for Quantum Science and Technology, University of Calgary, Calgary, AB T2N 1N4, Canada} 
\author{Mohammad Sahnawaz Alam}
\affiliation{Institute of Theoretical Physics, Wrocław University of Science and Technology, 50-370 Wrocław, Poland}
\author{Roberta Ramponi}
\affiliation{Department of Physics, Politecnico di Milano, Piazza Leonardo da Vinci, 32, 20133 Milano, Italy}
\affiliation{Institute for Photonics and Nanotechnologies (IFN-CNR), Piazza Leonardo da Vinci, 32, 20133 Milano, Italy}
\author{Paul E. Barclay}
\affiliation{Institute for Quantum Science and Technology, University of Calgary, Calgary, AB T2N 1N4, Canada}
\author{Andrea Chiappini}
\affiliation{Institute of Photonics and Nanotechnology of the National Research Council (IFN-CNR), Characterization and Development of Materials for Photonics and Optoelectronics (CSMFO) and The Centre for Materials and Microsystems (FBK-CMM), 38123 Trento, Italy} 
\author{Maurizio Ferrari}
\affiliation{Institute of Photonics and Nanotechnology of the National Research Council (IFN-CNR), Characterization and Development of Materials for Photonics and Optoelectronics (CSMFO) and The Centre for Materials and Microsystems (FBK-CMM), 38123 Trento, Italy} 
\author{Alexander Kubanek}
\affiliation{Institute for Quantum Optics, Ulm University, D-89081 Ulm, Germany}
\author{Angelo Bifone}
\affiliation{Department of Molecular Biotechnology and Health Sciences, University of Torino, Via Nizza 52, 10126 Torino, Italy}
\affiliation{Center for Sustainable Future Technologies, Istituto Italiano di Tecnologia, 10144 Torino, Italy} 

\author{Shane M. Eaton}
\affiliation{Institute for Photonics and Nanotechnologies (IFN-CNR), Piazza Leonardo da Vinci, 32, 20133 Milano, Italy}
\author{Anthony J. Bennett}
\affiliation{School of Engineering, Cardiff University, Queen’s Buildings, The Parade, Cardiff, CF24 3AA, United Kingdom}
\affiliation{Translational Research Hub, Maindy Road, Cardiff, CF24 4HQ, United Kingdom}

\email{BennettA19@cardiff.ac.uk}

\date{\today}

\begin{abstract}
Quantum emitters, such as the negatively charged nitrogen-vacancy center in diamond, are attractive for quantum technologies such as nano-sensing, quantum information processing, and as a non-classical light source. However, it is still challenging to position individual emitters in photonic structures whilst preserving the spin coherence properties of the defect. In this paper, we investigate single and ensemble waveguide-integrated nitrogen-vacancy centers in diamond fabricated by femtosecond laser writing followed by thermal annealing. Their spin coherence properties are systematically investigated and are shown to be comparable to native nitrogen-vacancy centers in diamond. This method paves the way for the fabrication of coherent spins integrated within photonic devices. 
\end{abstract}

\maketitle

\section{\label{Introduction} Introduction}

Optically addressable spins, including point defects in silicon carbide (SiC)\cite{Koehl2011RoomCarbide}, hexagonal boron nitride (hBN), and diamond\cite{Gottscholl2020InitializationTemperature,Barry2020SensitivityMagnetometry,Rondin2014MagnetometryDiamond}, are promising platforms for the realization of quantum sensing and quantum network applications\cite{Pompili2021RealizationQubits,Hensen2015Loophole-freeKilometres,Degen2017QuantumSensing}. In particular, negatively charged nitrogen-vacancy centers (NVs) in diamond have been extensively investigated and developed in various applications, such as high-resolution magnetic field sensing\cite{Balasubramanian2008NanoscaleConditions,Barry2020SensitivityMagnetometry}, and hybrid quantum networks\cite{Pompili2021RealizationQubits,Hensen2015Loophole-freeKilometres}. Although the spin coherence properties of NVs can reach up to microseconds at ambient temperature\cite{
Balasubramanian2009UltralongDiamond}, the relatively low brightness and low Debye-Waller factor hinder its application for quantum technologies\cite{Rondin2014MagnetometryDiamond}. Therefore, NVs are normally coupled into a photonic structure such as a solid immersion lens \cite{Hadden2010StronglyLenses}, nanopillar \cite{Babinec2010ASource,Huang2019AEmitters.}, optical waveguide\cite{LeSage2012EfficientWaveguide}, or cavity\cite{Riedel2017DeterministicDiamond} to improve their efficiency. However, due to diamond’s extreme hardness and chemical resistance, conventional methods such as plasma etching\cite{Babinec2010ASource}, ion implantation\cite{Eaton2019QuantumIrradiation}, or focused-ion beam\cite{Hadden2010StronglyLenses} have inevitably degraded their coherence and spectral properties. 

Recently, femtosecond laser writing has emerged as a versatile approach for the creation of photonic circuits\cite{Eaton2019QuantumIrradiation} with integrated quantum emitters. As femtosecond laser writing is a highly controllable fabrication process\cite{Chen2019LaserYield}, it is a promising method for scalably preparing quantum emitters in wide bandgap semiconductors, such as diamond\cite{Eaton2019QuantumIrradiation,Liu2013FabricationIllumination,Chen2017LaserDiamond,Fujiwara2023CreationIrradiation,Yurgens2021Low-Charge-NoiseLens}, hBN\cite{Hou2017LocalizedNitride,Gao2021FemtosecondNitride} and SiC\cite{Yang2023LaserTechnologies,Chen2019LaserCarbide}. In particular, laser-written NVs in diamond are of great interest due to the existence of mature and precise electron spin control protocols \cite{Barry2020SensitivityMagnetometry}. Compared with conventional approaches such as ion implantation\cite{Eaton2019QuantumIrradiation}, laser writing can place NVs deterministically \cite{Chen2017LaserDiamond,Salter2014ExploringCorrection} and within photonic structures inscribed in the bulk of diamond\cite{Hadden2018IntegratedWriting,Bharadwaj2019FemtosecondDiamond}. More recently, waveguide-integrated NVs (WGINVs) have been demonstrated by femtosecond laser writing\cite{Sotillo2016DiamondWriting,Alam2024DeterminingEnsembles, Hadden2018IntegratedWriting,Eaton2019QuantumIrradiation}. However, there have been few studies on their spin coherence\cite{Fujiwara2023CreationIrradiation,Stephen2019DeepCoherence}, and even fewer on WGINVs. These studies are crucial if the system is to be used for quantum applications such as quantum information processing or remote sensing. 

\begin{figure*}
\includegraphics[width=16.2cm]{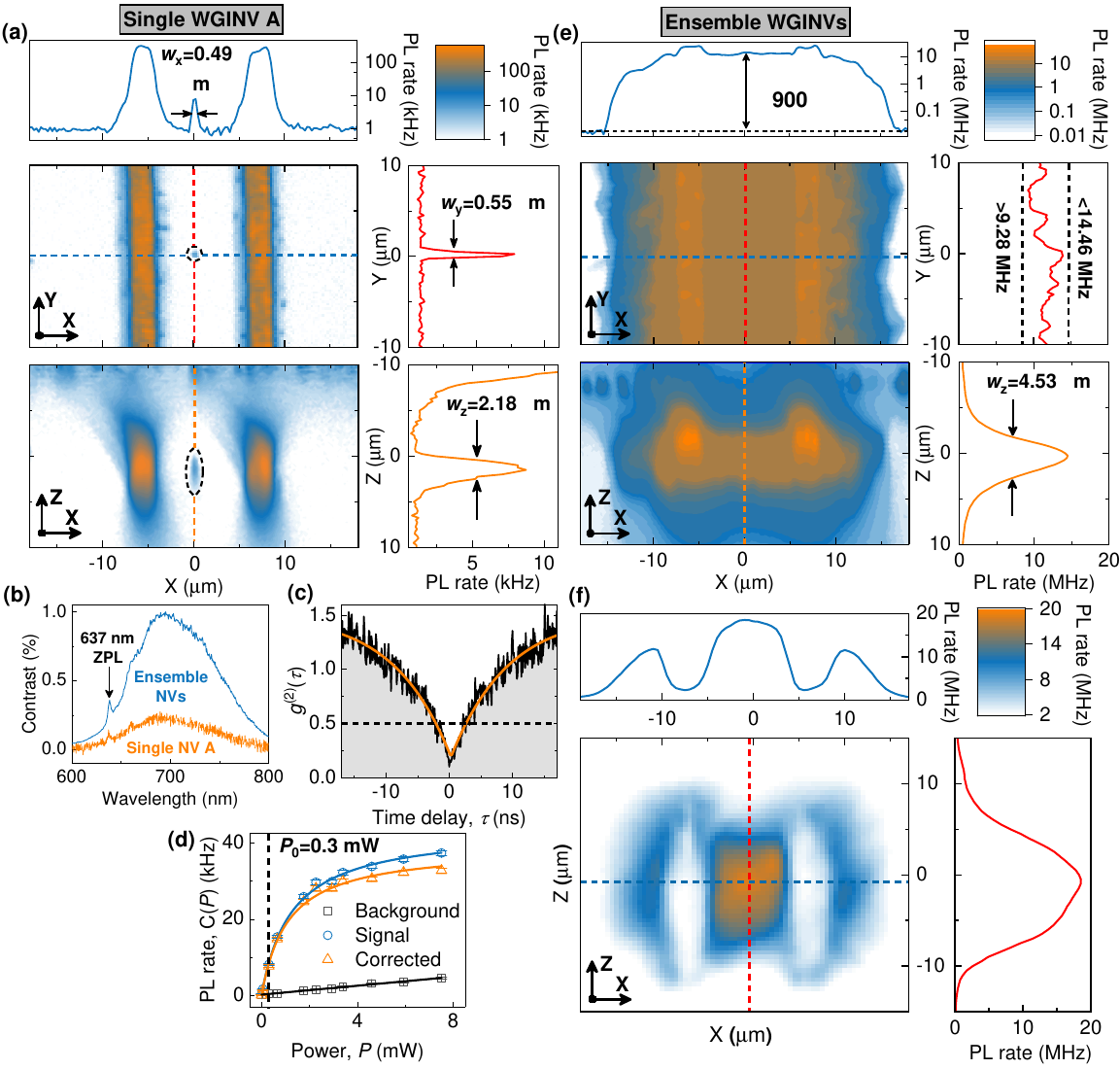}
\caption{Photoluminescence (PL) studies of laser-written WGINVs. (a) PL map of WGINV A in type IIa diamond, mapped from the z-direction (overhead) by scanning in x-y and x-z planes. The blue, red, and orange lines are slices along the dashed lines in (a), where $w_{x,y,z}$ are the Full Width at Half Maximum linewidth of $x,y,z$ direction Gaussian fitting curves for single WGINV A (see high-resolution PL map in Appendix \ref{Ensemble NVs density estimation calculation}). (b) The emission spectrum of single WGINV A and ensemble NVs. (c) The photon emission correlation of single WGINV A where the black line is raw data and the orange line represents the fitted curve (see Appendix \ref{Photon emission correlation}). (d) The power-dependent PL rate for single WGINV A. Dashed line is the test laser power $P_0$ used for density estimation of ensemble WGINVs. (e) PL map of WGINVs in type Ib diamond, mapped from the z-direction (overhead) by scanning in x-y and x-z planes. (f) PL map of the WGINVs in type Ib diamond mapped from the y-direction (along the waveguide).}
\label{Fig. 1}
\centering
\end{figure*}

In this paper, we characterize the spectral and spin coherence properties of single WGINVs in type IIa chemical vapor deposition (CVD) diamond and ensemble WGINVs in type Ib high-pressure high-temperature (HPHT) diamond, which are fabricated by femtosecond laser writing and subsequent thermal annealing. The home-built confocal optically detected magnetic resonance (ODMR) setup described in  Appendix \ref{ODMR setup} is used to coherently manipulate WGINVs, and characterize their spin coherence properties including the inhomogeneous dephasing time $T_2^*$, spin transverse relaxation time $T_2$ and longitudinal relaxation times $T_1$, via standard measurement protocols \cite{Barry2020SensitivityMagnetometry,Rondin2014MagnetometryDiamond}. Single WGINVs in type IIa diamond show excellent single-photon purity, and spin coherence properties comparable to native NVs. Additionally, we demonstrate coherent coupling with nuclear spins which could be used as a register for a hybrid quantum photonic circuit. In the type Ib diamond, we demonstrate the creation of ensemble WGINVs with up to 900 times intensity enhancement compared to the pristine diamond, as a result of increased NV density. This all-optical fabrication process therefore provides a cost-effective way to prepare ensemble NVs and realize photonically-integrated quantum sensing devices based on economical type Ib diamond with sub \SI{26}{\nano T\cdot\hertz^{-1/2}} DC magnetic field sensitivity. 

\section{\label{Results and discussion} Results and discussion}
\subsection{\label{Femtosecond laser written WGINVs} Femtosecond laser written WGINVs in diamond}

The WGINVs were created by femtosecond laser writing and subsequent annealing (see details in Appendix \ref{Sample annealing}). Parallel modification lines with \SI{13}{\micro\metre} separation were written to define the optical waveguides\cite{Eaton2019QuantumIrradiation}. In the case of nitrogen-rich type Ib diamond samples, thermal annealing drives the diffusion of vacancies that were generated within the laser-modified sidewalls defining the optical waveguides. These vacancies are captured by substitutional nitrogen atoms to form ensemble WGINVs. Conversely, for lower nitrogen content type IIa diamond, additional low energy static exposures were used to form vacancies localized near the center of within the approximately cubic micrometer-sized laser focal volume, and well aligned with the center of the previously laser-formed waveguides. Due to the lower background nitrogen, thermal annealing causes the diffusion of these vacancies which can yield single NVs near the center of the waveguide cross-section. In our case, we used two static exposures to produce single NVs at different longitudinal positions along the waveguide, which we labeled single WGINV A and WGINV B. \cite{Hadden2018IntegratedWriting}

\subsection{\label{PL study on WGINVs} PL study on WGINVs}

In Fig.~\ref{Fig. 1}(a), the two bright strips are the waveguide modification lines which host a high density of laser-written vacancies\cite{Eaton2019QuantumIrradiation}. Static single pulse exposures were observed to have formed isolated single WGINVs in the center of the optical waveguide in type IIa diamond, as shown in the top-down and cross-sectional PL map in  Fig.~\ref{Fig. 1}(a), where $w_{x,y,z}$ are the Full Width at Half Maximum linewidth of $x,y,z$ direction Gaussian fitting curves for single WGINV A and also represents our setup's optical resolution. A high-resolution PL map of the single WGINV A is shown in Appendix \ref{Ensemble NVs density estimation calculation}. The creation of NVs was confirmed by their PL emission spectrum with a \SI{637}{\nano\metre} zero-phonon line (ZPL) in Fig.~\ref{Fig. 1}(b). Meanwhile, the $g^{(2)}(0)$ < 0.2 in Fig.~\ref{Fig. 1}(c) and laser power-dependent PL saturation in Fig.~\ref{Fig. 1}(d)  indicate the quantized nature of its emission.

As explained above,  high-density ensembles of WGINVs are produced in type Ib diamond (Fig.~\ref{Fig. 1}(e)). We observe up to 900 times enhanced NV intensity in the waveguide relative to the pristine area in Fig.~\ref{Fig. 1}(e), revealing a corresponding NV density enhancement. Furthermore,  the ensemble WGINVs density $\rho$ is estimated to be \SI{14}{}-\SI{22}{ppb} in the waveguide region by taking the $C$\textsubscript{ensemble} of \SI{9.28}{}-\SI{14.46}{MHz} in the y plot in Fig.~\ref{Fig. 1}(e) and $C$\textsubscript{single} of \SI{7.7}{kHz} in Fig.~\ref{Fig. 1}(d) under $P_0$ of \SI{0.3}{mW}. The detailed calculation for the ensemble NVs density estimation is in Appendix \ref{Ensemble NVs density estimation calculation}. The z slice plot of Fig.~\ref{Fig. 1}(e) displays a Gaussian variation of PL rate, at the same depth as the modification lines. Furthermore, its width of \SI{4.53}{\micro\metre} is just a factor of 2 greater than the single WGINVs in Fig.~\ref{Fig. 1}(a) which is itself limited by the Rayleigh range of the microscope. This indicates that the laser-formed WGINVs are well-confined in the core of the waveguide region, where we investigate the spin coherence properties of those ensemble WGINVs in  Sec. \ref{Coherence characterization of WGINVs}.

To test the coupling between the ensemble WGINVs and waveguide in type Ib diamond, we scan the waveguide with excitation and collection along the waveguide direction, as shown in Fig.~\ref{Fig. 1}(f). Interestingly, we observe a stronger PL intensity in the waveguide region than in any other region which is different from the cross-section PL map in Fig.~\ref{Fig. 1}(e). This suggests that the pump laser and PL intensity in the waveguide core are guided, being efficiently coupled into the optics enhancing the collected PL intensity.

\begin{figure*}
\includegraphics[width=16.5cm]{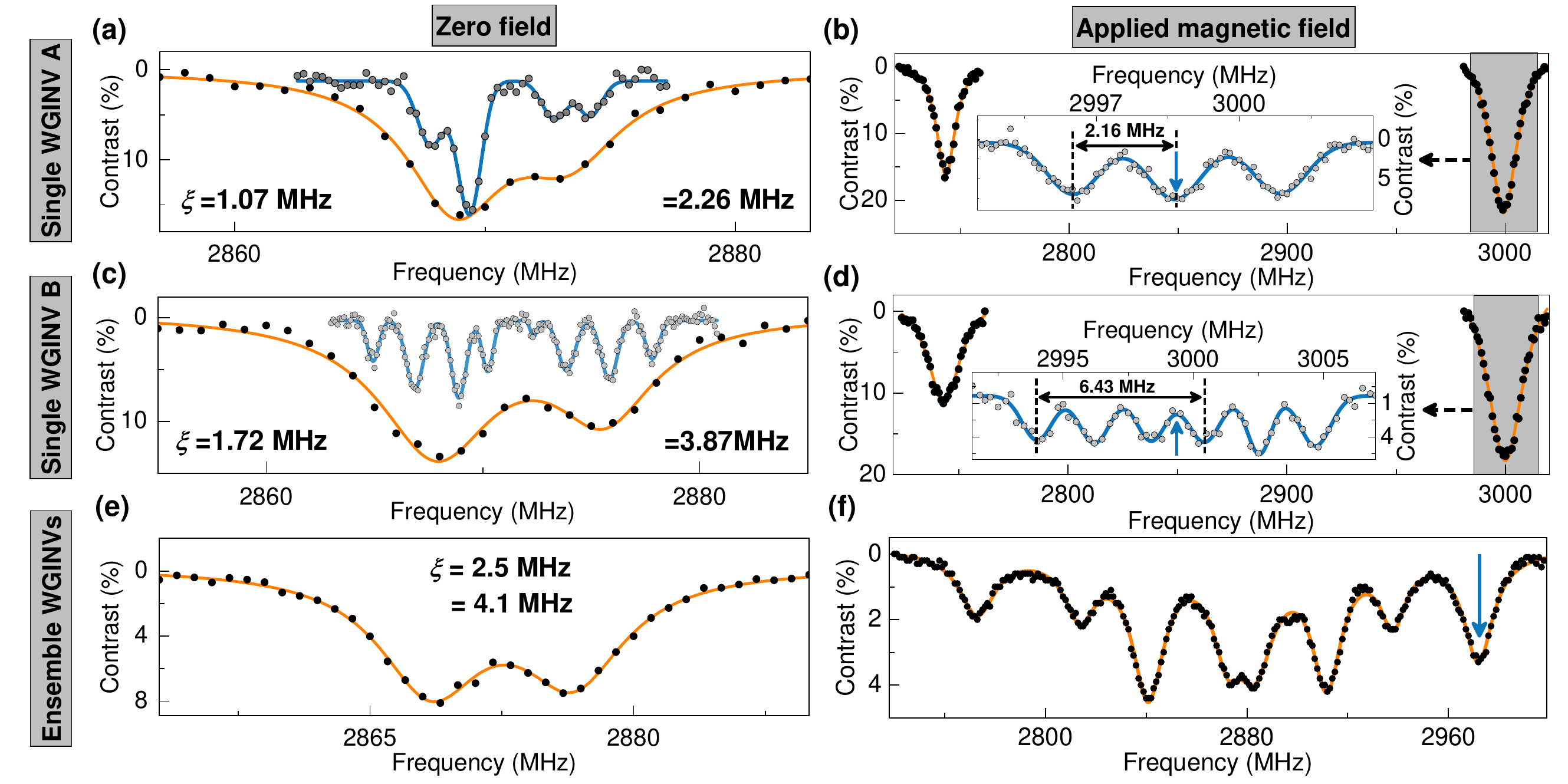}
\caption{ODMR spectra of single WGINVs (first and second rows) and ensemble WGINVs (third row). (a), (c) and (e) are the zero-field ODMR and (b), (d) and (f) are the ODMR spectrum with the applied magnetic field. The black (grey) scatter points and orange (blue) lines are the raw data and corresponding fitted curves in the CW (pulsed) model ODMR. The inserts of (b) and (d) are the hyperfine structure of the grey regions in (b) and (d), respectively. Blue arrows represent the target frequencies used for the time domain ODMR measurements.}
\label{Fig. 2}
\centering
\end{figure*}

\begin{figure*}
\includegraphics[width=16.5cm]{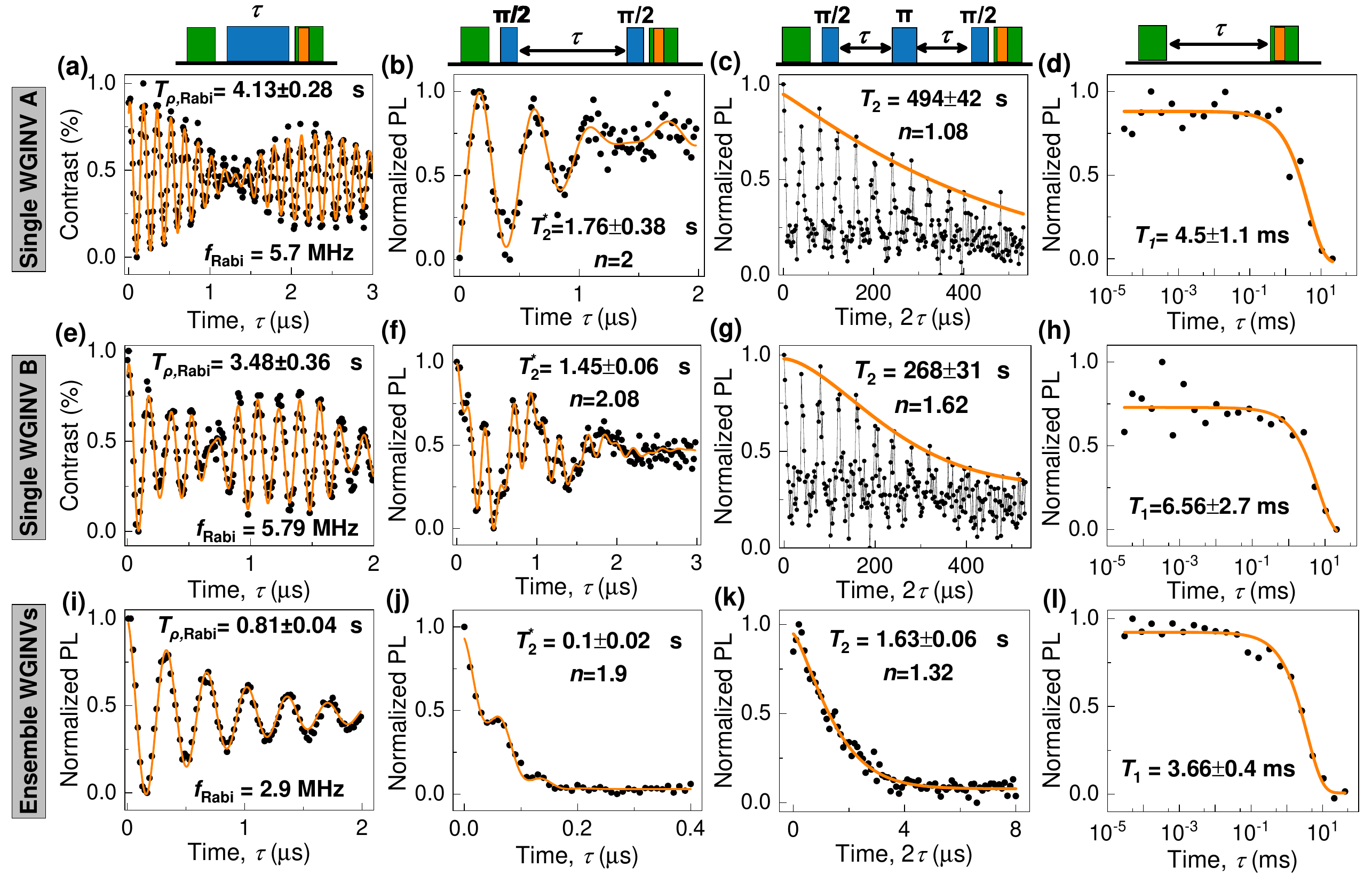}
\caption{Time domain ODMR spectrum of single WGINVs in type IIa diamond (from (a) to (h), first and second row) and ensemble WGINVs in type Ib diamond (from (i) to (l), third row). The first second, third, and fourth columns are Rabi, free induced decay (FID), Hahn echo, and relaxometry measurements, respectively. Meanwhile, their quantum control protocol is placed on the top of each column where the green, blue, and orange boxes represent laser and MW pulses, and the single photon detector gating window. The red curves are the fits to the raw data (black points). The envelopes of Rabi oscillation, FID, and Hahn data are fitted by equation,  $e^{-\left(\frac{t}{T_i}\right)^n}$, to extract $T_{\rho,Rabi}$, $T_2^*$ and $T_2$, where $T_i$ is the corresponding coherence time, and  $n$ is the stretched  exponent\cite{Wang2013SpinDiamond,Barry2020SensitivityMagnetometry,Liu2012ControllableTemperature}. The $T_1$ is obtained by fitting the single exponent decay equation.}
\label{Fig. 3}
\centering
\end{figure*}

\subsection{\label{Coherence characterization of WGINVs}Coherence characterization of WGINVs}

In this part, we systematically study the spin coherence properties of WGINVs in the frequency and time domains. A single NV's ground state is governed by the Hamiltonian\cite{Barry2020SensitivityMagnetometry},
\begin{equation}
    H = H_0+H_{S|E}+H_n
    \label{Hamiltonian}
\end{equation}
where $H_0$ consists of zero-field splitting parameter, \textit{D} $\sim$\SI{2.87}{GHz}, from the spin-spin interaction and Zeeman splitting $\gamma \: \textbf{\textit{B}}\cdot \textbf{\textit{S}}$ arising from the interaction between electron spin and external magnetic field, $H_{S|E}$ describes the electron spin interaction with local strain and electric fields \cite{Barry2020SensitivityMagnetometry}, and $H_n= \textbf{\textit{A}}\cdot\textbf{\textit{S}}\cdot \textbf{\textit{I}}$  is the interaction arising from electron spin and nuclear spin. The $\textbf{\textit{S}}$ and $\textbf{\textit{I}}$ are electron and nuclear spin operators, the $\gamma$ is the gyromagnetic ratio, and $\textbf{\textit{A}}$ is the nuclear hyperfine tensor. The diagonalization of the NV ground state Hamiltonian yields an approximation of the spin resonance ($m_s=\pm1$) frequencies: \cite{Hilberer2023EnablingGPa} 
\begin{equation}
    v_{\pm}=D+\xi \pm 2\Delta
    \label{ODMR resonance frequency}
\end{equation}
where the $\xi$ describes the shift component which may be affected by temperature, transverse magnetic field, axial or hydrostatic strain, and axial electric fields. Meanwhile, $\Delta$ describes the splitting component which could originate from axial magnetic fields, non-axial or anisotropic strain, non-axial electric fields, and the nuclear hyperfine interaction. For ensembles of NV centres, the effect of these external fields must be considered for each of the four possible orientations of NV centre\cite{Barry2020SensitivityMagnetometry}.

\subsubsection{\label{Frequency domain ODMR spectrum} Frequency domain ODMR spectrum}

Overall, single WGINV A and single WGINV B in the type IIa diamond and ensemble WGINVs in the type Ib diamond display a similar shape of zero-field continuous wave (CW) ODMR with three consistent features (first column of Fig.~\ref{Fig. 2}). First, they all show positive shift $\xi$ of a few \SI{}{MHz} compared to $D$ =\SI{2.87}{GHz}. We attribute this to the compressive strain arising from laser-written modification lines providing a component of strain along the axis of the NV centre\cite{Doherty2014ElectronicPressure,Eaton2019QuantumIrradiation}. Second, the lower resonance frequency exhibits slightly higher ODMR contrast due to the excitation with a linearly polarized microwave (MW) magnetic field on a mixed sublevel $m_s=\pm1$ state split by non-axial or anisotropic strain\cite{Hilberer2023EnablingGPa,Eaton2019QuantumIrradiation}. Third,  two peaks are observed with splitting $\Delta$ of a few \SI{}{MHz}. Using a pulsed ODMR sequence the resolution of the experiment is greatly improved, leading to more structure being resolved for both single WGINVs. For single WGINV A, the splitting $\Delta$ of \SI{2.26}{\mega\hertz} in CW ODMR and four resolvable resonance peaks in pulsed ODMR indicates the presence of nuclear \ce{^{14}N} interaction and non-axial or anisotropic strain in the waveguide region\cite{Zheng2019Zero-FieldDiamond, Hilberer2023EnablingGPa, Mittiga2018ImagingDiamond}.  For single WGINV B, the splitting $\Delta$ of \SI{3.87}{\mega\hertz} in CW ODMR and 8 resolvable resonance peaks in pulsed ODMR implies that a complex combination of the non-axial or anisotropic strain and strong coupled nuclear \ce{^{14}N} and \ce{^{13}C} leads to splitting\cite{Wang2022Zero-fieldSurfaces, Zheng2019Zero-FieldDiamond}.

For the ensemble WGINVs in type Ib diamond we observe a positive shift $\xi$ of \SI{2.5}{\mega\hertz}, and a zero-field splitting $\Delta$ of \SI{4.1}{\mega\hertz}. We note that in this regime the contributions from each of the four orientations of the NV centre ensemble cannot be individually resolved. However, the splitting, and shift of the combined signal are consistent with the expected compressive strain arising from the laser-written modification lines \cite{Alam2024DeterminingEnsembles,Sotillo2016DiamondWriting}, as well as some additional effects from the local electric field from over 200 ppm charge impurities \cite{Mittiga2018ImagingDiamond}. 

To further lift the degeneracy of $m_{s}$ = $\pm 1$ and quantify the hyperfine interaction strength between NV electron spin and nearby strong coupling nuclear spins, a \SI{4.6}{mT} magnetic field is applied for the single WGINVs in type IIa diamond. The typical \SI{2.16}{\mega\hertz} \ce{^{14}N} hyperfine splitting observed in pulsed ODMR\cite{Barry2020SensitivityMagnetometry} of single WGINV A  is shown in grey data points in Fig.~\ref{Fig. 2}(b). For single WGINV B,  we observed six anti-peaks in the insert of Fig.~\ref{Fig. 2}(d) revealing the \SI{6.43}{\mega\hertz} \ce{^{13}C} coupling strength\cite{Dreau2012High-resolutionDiamond} in combination with \SI{2.16}{\mega\hertz} \ce{^{14}N} hyperfine splitting. The fact that the nuclear hyperfine splitting is resolvable in both zero field and non-zero magnetic field ODMR, highlights the excellent coherence properties of NVs and their potential as quantum registers. For ensemble WGINVs in Ib, in Fig.~\ref{Fig. 2}(f), a \SI{\sim5.17}{mT} magnetic field is applied to lift the degeneracy of the  $m_{s}$ = $\pm 1$ transitions along the four NV orientations. The central frequencies marked with blue arrows in Figs.~\ref{Fig. 2}(b), (d), and (f) are now used for time domain measurements in Sec. \ref{Time domain ODMR spectrum}. 

Overeall, we conclude that these WGINVs are sensitive enough to probe the microscopic strain\cite{Eaton2019QuantumIrradiation}, the electric field, and the nuclear spin environment\cite{Wang2022Zero-fieldSurfaces, Zheng2019Zero-FieldDiamond}. 

\subsubsection{\label{Time domain ODMR spectrum} Time domain ODMR spectrum}

To get a deeper understanding of the coherence properties we have measured the $T_2^*$, $T_2$, and $T_1$ of WGINVs with standard protocols, shown at the top of Fig.~\ref{Fig. 3}. Due to the existence of strongly coupled nuclear spins the dynamics of the electron in single NV A and NV B are dominated by these hyperfine coupling frequencies. This results in beats being observed in Figs.~\ref{Fig. 3}(a) and (e) for single WGINV A and single WGINV B with Rabi frequency $f\textsubscript{Rabi}$ around \SI{5.7}{\mega\hertz} and Rabi dephasing time $T_{\rho,}\textsubscript{Rabi}$ around \SI{4}{\micro\second}. The Rabi dephasing time  $T_{\rho,}\textsubscript{Rabi}$ could be extended to over \SI{10}{\micro\second} by reducing the MW power broadening with a  Rabi frequency under \SI{1}{MHz} (see Appendix \ref{Rabi oscillations at weak MW power}.)  In comparison, a Rabi frequency of \SI{2.9}{\mega\hertz} coherently drives ensemble WGINVs in type Ib diamond with $T_{\rho,}\textsubscript{Rabi}$ of \SI{0.81}{\micro\second} in Fig.~\ref{Fig. 3}(i). These Rabi frequencies are used for the following multi-pulse experiments.

The $T_2^*$ of single WGINV A and WGINV B are around \SI{1.5}{\micro\second} from free induction decay (FID) measurements as shown in Figs.~\ref{Fig. 3}(b) and (f), which is limited by the naturally abundant isotopic \ce{^{13}C} bath\cite{Rondin2014MagnetometryDiamond,Barry2020SensitivityMagnetometry}. Moreover, there is one dominant oscillation of \SI{2.16}{\mega\hertz} for single WGINV A in Fig.~\ref{Fig. 3}(b) and three frequencies of oscillation (\SI{1.07}{\mega\hertz}, \SI{3.23}{\mega\hertz} and \SI{5.4}{\mega\hertz}) for single WGINV B in Fig.~\ref{Fig. 3}(f) which are clearly resolved in FID curves. This is consistent with the hyperfine structure of \ce{^{14}N} and strong coupling \ce{^{13}C} in Figs.~\ref{Fig. 2}(b) and (d). This indicates their excellent spin dephasing time and strong interaction between these spins. Regarding spin transverse relaxation time, both single WGINV A and single WGINV B exhibit the $T_2$ over \SI{268}{\micro\second} which is comparable with native single NV in type IIa diamond. The collapse and revival behavior in both single WGINV A and WGINV B match the spin bath Larmor precession frequency (\SI{50}{k\hertz}) in the Hahn echo curve, which indicates that the \ce{^{13}C} spin bath is the dominant decoherence source.  Moreover, we highlight that for single WGINV A,  $T_2$ of \SI{494}{\micro\second} is longer than most of NVs integrated into photonics structures\cite{Li2015CoherentDiamond,Andrich2014EngineeredFluid,Neu2014PhotonicDiamond}. Although single WGINV B has a slightly shorter $T_2$ of \SI{268}{\micro\second}, the interaction of electron and strong coupling \ce{^{13}C} nuclear spin could be used as a quantum memory register\cite{Bradley2019AMinute,Robledo2011High-fidelityRegister}. Finally, in Figs.~\ref{Fig. 3}(d) and (h), both single WGINV A and WGINV B display $T_1$ over \SI{4.5}{ms} as the native single NV in type IIa diamond\cite{Jarmola2012Temperature-Diamond}. 

Although the \SI{0.1}{\micro\second} $T_2^*$ of ensemble WGINVs is nearly 10 times shorter than single WGINVs due to its inhomogeneous electron spin environment (Fig.~\ref{Fig. 2}(j)), the value of its coherence time is still consistent with those of native NVs in type Ib diamond\cite{Rondin2014MagnetometryDiamond}. The transverse magnetic field has an impact on the ensemble NVs\cite{Stanwix2010CoherenceDiamond}, a  $T_2$ of \SI{1.63}{\micro\second} is obtained which is comparable with singe NV's coherence properties in type Ib diamond, and mainly limited by the nitrogen electronic spin bath (P1 centers)\cite{Barry2020SensitivityMagnetometry}. Moreover, the $T_1$ curves of ensemble NVs do not show any obvious decay within \SI{1}{ms}, which is comparable with the relaxometry performance of native NVs\cite{Jarmola2012Temperature-Diamond,Rondin2014MagnetometryDiamond}. Therefore, it indeed proves that the laser writing process produces ensemble WGINVs in the waveguide region without compromising their spin coherence properties. 
\subsubsection{\label{Sensitivity} Sensitivity }

By taking the PL rate $C\sim$\SI{30}{kHz}, ODMR contrast $\Lambda\sim20\%$, readout duration time $t_{L}\sim$\SI{0.5}{\micro\second}, $T_2^*$ of \SI{1.76}{\micro\second}, and $T_2$ of \SI{494}{\micro\second} for single WGINV A, its photon-shot-noise-limited DC ($\eta$\textsubscript{dc}) and AC ($\eta$\textsubscript{ac}) magnetic field sensitivity are estimated around \SI{174.9}{\nano T\cdot\hertz^{-1/2}} and \SI{10.4}{\nano T\cdot\hertz^{-1/2}}, in the pulsed mode measurement by equations\cite{Rondin2014MagnetometryDiamond},
\begin{equation}
\eta\textsubscript{dc}\sim\frac{\hbar}{g\mu_B}\frac{1}{\Lambda\sqrt{Ct_L}}\times\frac{1}{\sqrt{T_2^*}}
\label{Eq(6)}
\end{equation}
\begin{equation}
\eta\textsubscript{ac}=\eta\textsubscript{dc}\sqrt{\frac{T_2^*}{T_2}}
\label{Eq(7)},
\end{equation}
respectively. The $g\sim$ 2\textit{.}0 is the Landé $g$-factor, $\mu_{B}$ is the Bohr magneton, and $\hbar$ is the reduced Planck constant. 

For ensemble WGINVs in type Ib diamond, the $\eta$\textsubscript{dc} and $\eta$\textsubscript{ac} are estimated around  \SI{25.7}{\nano T\cdot\hertz^{-1/2}} and \SI{6.6}{\nano T\cdot\hertz^{-1/2}}, by taking readout duration time $t_{L}\sim$\SI{0.5}{\micro\second}, $T_2^*$ of \SI{0.1}{\micro\second}, $T_2$ of \SI{1.53}{\micro\second}, the PL rate $C\textsubscript{ensemble}\sim $\SI{0.9}{GHz}, and ODMR contrast $\Lambda\sim3.3\%$ for the external peak (marked by blue arrow) in Fig.~\ref{Fig. 2} (f). We note that the PL rate of the ensemble WGINVs used here was not at saturation. Therefore, the sensitivity could be improved by increased laser power excitation, or by increasing the number of NV centers probed through excitation and collection via the waveguide mode.

\section{\label{Conclusion and outlook} Conclusion and outlook}

We fabricated and characterized laser-written coherent single WGINVs in type IIa diamond and ensemble WGINVs in type Ib diamond. The density of ensemble WGINVs in the type Ib diamond sample was estimated to be 14-22 ppb. The spin coherence properties of the WGINVs were similar to that of the native NVs as evidenced by the summary in Table \ref{tb1:Sample information}, further demonstrating the promise of laser writing to realize an integrated photonics platform incorporating quantum emitters. 
\begin{table}[ht]
  \caption{Comparison of native NVs from L. Rondin et al's \cite{Rondin2014MagnetometryDiamond} and our work.}
    \label{tb1:Sample information}
  \begin{tabular}{lllll}
    \hline
 Diamond type&  IIa native&IIa WGINV A&Ib native&Ib WGINVs\\
 \hline
 Synthesis&  CVD&CVD&HPHT&HPHT\\
 N& < 5 ppb&< 5 ppb& 200 ppm& 200 ppm\\
 NV& Single&Single&Single&14-22 ppb\\
 $T_2^* \left(\SI{}{\micro\second}\right)$&3&1.76& 0.1&0.1\\
 $T_2\left(\SI{}{\micro\second}\right)$ &300&494&1&1.63\\
 $\eta\textsubscript{dc}\left(\SI{}{\nano T\cdot\hertz^{-1/2}}\right)$&300 &174.9& 1500&25.7\\
 $\eta\textsubscript{ac}\left(\SI{}{\nano T\cdot\hertz^{-1/2}}\right)$&30&10.4&500 &6.6\\
     \hline
 \end{tabular}
 \end{table}
 
 These highly coherent single WGINVs couple to nearby nuclear spins with excellent coherence which could be used as a quantum register within photonic integrated circuits. In terms of ensemble WGINVs, this all-optical fabrication technique paves the way for a cost-effective waveguide-integrated quantum sensing device based on the more economical type Ib diamond. Future work will focus on the optimization of the NV creation process, including the annealing and laser writing processes, to further increase the NV density and reduce the non-NV spin noise. 

\begin{acknowledgments}
The authors acknowledge financial support provided by EPSRC via Grant No. EP/T017813/1 and EP/03982X/1 and the European Union's H2020 Marie Curie ITN project LasIonDef (GA No. 956387). IFN-CNR is thankful for support from the projects QuantDia (FISR2019-05178) and PNRR PE0000023 NQSTI funded by MUR (Ministero dell'Università e della Ricerca). V. B. acknowledges the support of the Alexander von Humboldt Foundation.
\end{acknowledgments}

\section*{Author Contributions}
The corresponding author identified the following author contributions, using the CRediT Contributor Roles Taxonomy standard:

Conceptualization: YG, JPH, AJB, SME;
Methodology: YG, JPH, AJB, SME;
Software: YG;
Investigation: YG, VB;
Validation: FG, GC, VKK;
Resources: AB, SME;
Writing - Original Draft: YG;
Writing - Review \& Editing: All;
Supervision: JPH, RR, PEB, AC, MF, AK, AB, SME, AJB;
Funding Acquisition: JPH, VB, RR, AK, AB, SME, AJB.

\section*{Data Availability Statement}
Data supporting the findings of this study are available in the Cardiff University Research Portal at http://doi.org/xx.xxxx
\nocite{*}
\appendix

\section{\label{ODMR setup}ODMR setup}

A CW \SI{532}{\nano m} laser (Crystal Laser) was modulated by an acoustic-optic modulator (ISOMET 553F-2) with < \SI{10}{\nano s} rise and fall time. A 2-axis Galvo mirror (GVS002) and NA=0.9 objective were integrated into a 4f imaging system for 2D x-y scanning. Depth scanning (z) was implemented by a motorized sample stage. The PL was optically filtered by the dichroic mirror, \SI{532}{\nano m} long-pass filter, and \SI{650}{\nano m} long-pass filter, before detection on SPCM-AQRH silicon avalanche photodiodes (Excelitas) or a spectrometer with a silicon CCD. The optional ND filter is also used to keep the PL rate within the APD working range (\SI{2}{\mega\hertz}). The microwave (MW) field is generated by an E4438B MW source, modulated by a RF switch (ZASWA-2-50DRA+) and amplified by a MW amplifier (ZHL-42W+), eventually transmitted to the sample by a patch antenna. The Rabi oscillation, free induction decay, Hahn echo and $T_{1}$ measurements are implemented by the standard protocol\cite{Rondin2014MagnetometryDiamond}.

\section{\label{Sample annealing}Laser writing and annealing}

A commercial femtosecond laser (Menlo Systems BlueCut) that produces linearly polarized pulses with a wavelength of \SI{515}{\nano\metre} (second harmonic), repetition rate of 500 kHz, and a duration of \SI{300}{\femto\second} was used for femtosecond laser writing. The laser pulse was focused using a high numerical aperture oil immersion objective lens (NA = 1.25) beneath the surface of synthetic diamond samples to write waveguides and NVs. The bulk diamond samples were mounted on precision translation stages for three-dimensional control. The pulse energy was controlled using a combination of a motorized half-wave plate and a fixed linear polarizer. The laser writing fabrication parameters are shown in Table \ref{tbl:notes1}. 
\begin{table*}[ht]
  \caption{Experimental parameters for laser fabrication}
    \label{tbl:notes1}
  \begin{tabular}{lllll}
    \hline
   Diamond & Laser exposure&  Pulse energy (nJ)& Type II spacing (\SI{}{\micro\metre})& Depth (\SI{}{\micro\metre})\\
    \hline
    IIa (WG)  &\SI{500}{\kilo\hertz}&  60&13&35.0\\
 IIa (NVs)&single pulse& 28& N/A&22.5\\
    Ib (WG \& NVs)&\SI{500}{\kilo\hertz}& 100&13&25.0\\
 \hline
 \end{tabular}

\end{table*}

Annealing of the diamond samples was performed in a tubular horizontal furnace Lenton LTF15/50/450. The samples were placed in a quartz boat and covered with diamond grit to protect the surface. In order to purge the furnace chamber, oxygen was extracted using a diaphragm pump and a nitrogen flow for 1 h. Thermal treatments were carried out in a nitrogen atmosphere following three steps: first, the temperature was ramped from room temperature to 1000$^{\circ}$C over the course of 3 hours and then kept at 1000$^{\circ}$C for another 3 hours, and the furnace was finally switched off and allowed to cool down to room temperature. 

\section{\label{Ensemble NVs density estimation calculation} Ensemble NVs density estimation calculation}

\begin{figure}[ht]
    \centering
    \includegraphics[width=5.5cm]{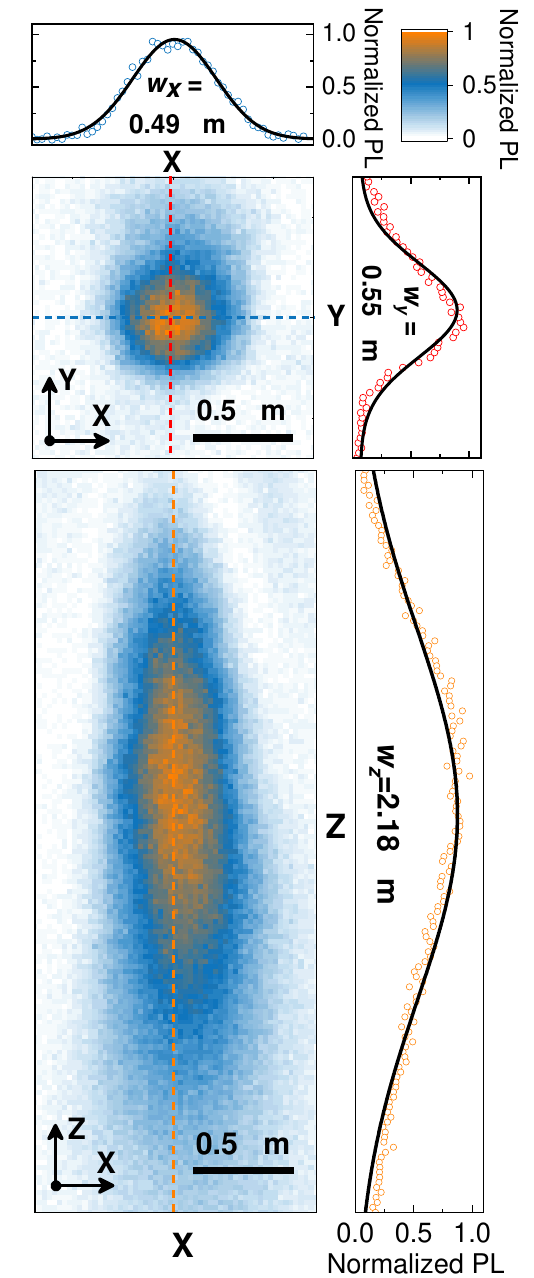}
    \caption{High-resolution confocal normalized PL image of single WGINVs A in type IIa diamond where the blue, red, and orange data in the top and right plates are x, y, and z slices along the dashed lines in confocal PL image.}
    \label{Fig. S1}
\end{figure}

To estimate the ensemble WGINVs density in type Ib diamond, we compare a single WGINV A's PL intensity at power $P_0$ = \SI{0.3}{mW}, which is much lower than its saturation power \SI{1.10}{mW} in Fig.~\ref{Fig. 1} (d). In this regime, the detected photon rate $C$\textsubscript{single}($P$) linearly scales with the laser power, and is given by:
\begin{equation}
      C\textsubscript{single}(P)=KP
      \label{Eq(1)}
\end{equation}
where $K$ is the linear scaling ratio between PL rate and laser power. 
We assume the ensemble of NVs in type Ib diamond to be inhomogeneously pumped due to the spatial Gaussian distribution of laser power.  Therefore, we obtain the equation
\begin{subequations}
\begin{eqnarray}
C\textsubscript{gaussian}(P_0)&=& \iiint_V KP_0G(x)G(y)G(z) \rho dV, \label{Eq. 3(b)}
\\
C\textsubscript{gaussian}(P_0)&=&C\textsubscript{single}(P_0)\rho A_x A_y A_z\label{Eq. 3(c)}
\end{eqnarray}
\end{subequations}

where the $G(x)$, $G(y)$, and $G(z)$ are the Gaussians fitted to the slices along the \textit{\textit{x}, y} and \textit{z} in Fig.~\ref{Fig. S1}, $A_{x,y,z}$ are their corresponding integrals of the Gaussian fitting curves in Fig.~\ref{Fig. S1}. The ensemble NVs density $\rho$ is estimated to be 14- 22 ppb in the waveguide region by taking the $C$\textsubscript{ensemble} of \SI{9.28}{}-\SI{14.46}{MHz} and $C$\textsubscript{single} of \SI{7.7}{kHz} under $P_0$ of \SI{0.3}{mW}. We assume the light is collected with equal intensity from all 4 allowed NV orientations in the ensemble. 
NVs density is also estimated at $\rho$ = 0.024 ppb in the pristine region by taking the $C$\textsubscript{ensemble} of \SI{16.1}{kHz}. 

In contrast, we note that a more simplistic model of NV centre ensemble excitation which assumes the density to be uniform, could lead to an underestimate of NV density. For example, if we compare to a model which uses uniform excitation and collection from a spheroid with dimensions $w_{x,y,z}$ of the full width at half maximum linewidth of the confocal microscope's point spread function, we can obtain 
\begin{subequations}
\begin{eqnarray}
C\textsubscript{uniform}(P_0)&=& \iiint_V KP_0 \rho dV, \label{Eq. C2a}
\\
C\textsubscript{uniform}(P_0)&=&C\textsubscript{single}(P_0)\rho \frac{4\pi}{3} w_xw_yw_z\label{Eq. C2b}
\end{eqnarray}
\end{subequations}
In this case, there is an additional factor of $\frac{4 \pi}{3}\sim4$ compared to the corresponding integration of the three-dimensional Gaussian distribution. Consideration of the non-uniform laser power distribution could also be used to study the laser-induced inhomogeneous charge and spin initialization dynamics of ensemble NVs, which is crucial for optimizing the ensemble-based quantum sensing protocol.

\section{\label{Photon emission correlation}Photon emission correlation}

Photon emission correlation was recorded using two detectors in a Hanbury-Brown and Twiss interferometer. The $g^{(2)}$(\(\tau\)) data is shown using the empirical equation,
\begin{equation}
\begin{split}
g^{(2)}(\tau)= & 1-C_1e^{-|\tau-\tau_0|/\tau_1}
+C_2e^{-|\tau-\tau_0|/ \tau_2}+C_3e^{-|\tau-\tau_0| /\tau_3} \
\label{eq:g2_empirical}
\end{split}
\end{equation}
Here, $\tau_0$ is the delay time offset of the two detectors, $\tau_{1}$ is the antibunching time, $C_{1}$ is the antibunching amplitude, $\tau_{i}$ for \emph{i} \(\geq\) 2 are bunching times, and $C_{i}$ for \emph{i} \(\geq\) 2 are the corresponding bunching amplitudes. The fitting parameters are shown in table \ref{tb2:g2fittingparameters}.

\begin{table}[ht]
  \caption{Fitting parameters of $g^{(2)}(\tau)$}
    \label{tb2:g2fittingparameters}
  \begin{tabular}{lll}
    \hline
   Parameters&  Value & Standard Error
\\
    \hline

    $\tau_0$ (ns)& 0.275&0.012
\\
 $C_1$& 1.481& 0.011
\\
 $\tau_1$ (ns)& 9.48 & 0.12
\\
 $C_2$& 0.365& 0.061
\\
 $\tau_2$ (ns)& 114 & 17
\\
 $C_3$& 0.313& 0.067
\\
 $\tau_3$ (ns)& 312 & 29
\\\hline
 \end{tabular}

\end{table}
\section{\label{Rabi oscillations at weak MW power}Rabi oscillations at weak MW power}
\begin{figure}[ht]
    \centering
    \includegraphics[width=8.5cm]{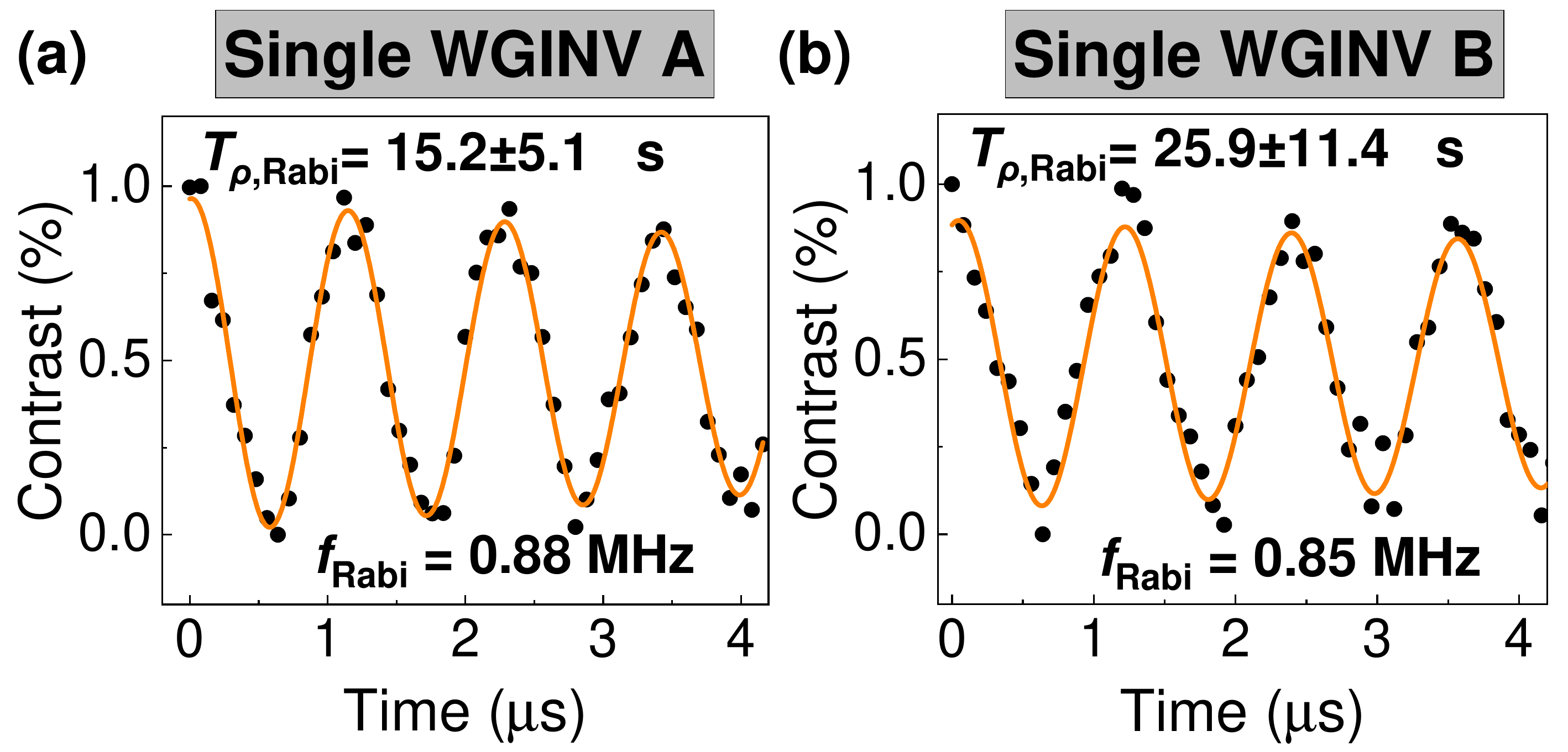}
    \caption{Rabi oscillations for single WGINV A (a) and single WGINV B (b) at weak MW power.}
    \label{Fig. S2}
\end{figure}

Figs.~\ref{Fig. S2} show the Rabi oscillation for single WGINV A and WGINV B at weak MW power, showing $T_{\rho, Rabi}$ over \SI{10}{\micro\second} with $f_{Rabi}$ less than \SI{1}{\mega\hertz}. Compared to Fig.~\ref{Fig. 3}(a), there is no beating due to the MW being only strong enough to drive one transition in Figs.~\ref{Fig. 2}(a) and (b).

\section*{Reference}
\bibliographystyle{ieeetr}
\bibliography{references}

\end{document}